\documentclass[aps,preprint,tightenlines,nofootinbib,showpacs]{revtex4}

\usepackage{graphicx}

\newcommand{\LQ}{L\!Q}
\newcommand{\LQb}{\overline{L\!Q}}
\newcommand{\mlq}{\mbox{$M_{\scriptstyle \LQ}$}}
\newcommand{\msq}{\mbox{$M^2_{\scriptstyle \LQ}$}}

\newcommand{\alps}{\mbox{$\alpha_{\mbox{\scriptsize s}}$}}
\newcommand{\alpsq}{\mbox{$\alpha_{\mbox{\scriptsize s}}^{2}$}}

\begin{document}

\preprint{CERN-PH-TH-2004-207, DESY 04-200, Edinburgh 2004/28, 
          FERMILAB-PUB-04-298-T, PSI-PR-04-12}

\title{Pair production of scalar leptoquarks at the LHC}

\author{M.~Kr\"amer}
\affiliation{School of Physics, The University of Edinburgh,
Edinburgh EH9 3JZ, UK
\footnote{present address: Institut f\"ur Theoretische Physik E,
RWTH Aachen, D-52056 Aachen, Germany.}}

\author{T.~Plehn}
\affiliation{Theory Division, CERN, 1211 Geneva 23, Switzerland\\
and\\ 
Max Planck Institut f\"ur Physik, 80805 M\"unchen, Germany}

\author{M.~Spira\footnote{Supported in part by the Swiss Bundesamt
f\"ur Bildung und Wissenschaft.}}
\affiliation{Paul Scherrer Institut PSI, CH-5232 Villigen PSI, Switzerland}

\author{P.M.~Zerwas}
\affiliation{Deutsches Elektronen-Synchrotron DESY
\footnote{Permanent address.}, D-22603 Hamburg, Germany\\
and\\ 
Fermi National Accelerator Laboratory, P.O.Box 500, Batavia IL 60510, USA}

\date{\today}

\begin{abstract} 
Theoretical predictions for the production cross sections of
leptoquarks at the CERN LHC are presented including higher-order QCD
corrections. These corrections reduce the dependence on the
renormalization/factorization scales significantly. Moreover, they are
required to exploit the leptoquark mass reach of the LHC
experiments. In this sequel to an earlier analysis performed for the
Tevatron collider we extend the leptoquark analysis to the LHC energy.
\end{abstract}


\maketitle

Leptoquarks \cite{Buchmuller:1986zs} have been searched for in the
past at all high energy colliders operating at the energy frontier. If
the Yukawa couplings for scalar leptoquark couplings to quarks and
leptons of the first and second generation [on which we will focus in
this report] is kept at small values, as required by low-energy
precision experiments~\cite{Kalinowski:1997fk}, the
most stringent model-independent bounds on leptoquark masses, for
branching ratios $BR = 1$ to charged leptons, have been set with $\mlq
\geq$ 230 GeV and 241 GeV by the FNAL Tevatron CDF experiment for
first and second generation leptoquarks, respectively, and
correspondingly with 238 and 186 GeV by the D0 experiment as reported
recently in Ref.~\cite{Wang:2004cj}; these recent individual values
come close to the earlier combined limit of 242 GeV for the first
generation in Ref.~\cite{Grosso-Pilcher:1998qt}. The D0 and CDF
experiments have also searched for leptoquarks decaying into quark and
neutrino final states, resulting in a lower mass limit of 95~GeV and
117~GeV, respectively, in these decay channels~\cite{Abazov:2001ic}.
Bounds of 290 GeV for the first generation could be set by the Zeus
\cite{Chekanov:2003af} and the H1 experiments \cite{Adloff:2001cp} in
direct electron-quark formation at DESY's HERA collider for Yukawa
couplings of size 0.1 and electromagnetic strength, respectively. A
limit of 300 GeV is expected to be reached finally also by the
Tevatron at the end of Run~II~\cite{Rolli}.

The search for these novel particles will be continued soon at the
CERN LHC.  Preliminary feasibility studies by the LHC experiments
ATLAS~\cite{ATLAS} and CMS~\cite{Abdullin:1999im} indicate that clear
signals can be established for masses up to about $\mlq \simeq$ 1.3
to 1.4 TeV for first and second generation scalar leptoquarks, with a
final reach of presumably 1.5 TeV.

In this Brief Report we present the cross section for pair production
of scalar leptoquarks at the LHC including next-to-leading order QCD
corrections.%
\footnote{Note that QCD corrections to the production of vector
  leptoquarks are not under proper theoretical control as vector
  leptoquarks are described by a non-renormalizable effective theory.}
The results are based on the calculation presented earlier in
Ref.~\cite{Kramer:1997hh} for Tevatron energies. The higher-order
corrections must be included in the theoretical predictions for the
production cross sections in order to reduce the strong dependence on
the renormalization and factorization scales which the lowest-order
Born term calculation is bedevilled with. Moreover, for the standard
choice of the common renormalization/factorization scale near the
leptoquark mass, the cross section is increased by the QCD corrections
and the mass range that can be probed experimentally, is extended
correspondingly.

\begin{figure}[hb]
\includegraphics[bb=50 525 550 725,width=11.0cm,clip=]{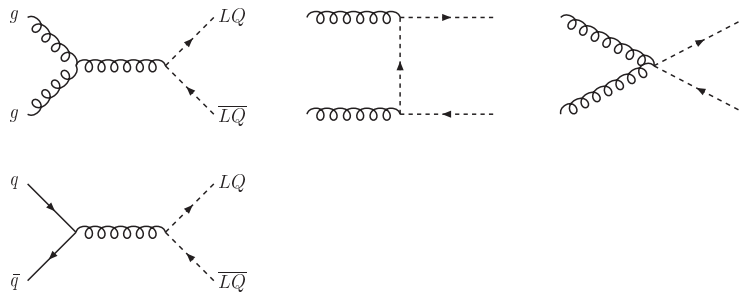}
\caption{Generic set of leading-order Feynman diagrams for leptoquark 
 pair production through gluon-gluon fusion and quark-antiquark
 annihilation.}
\label{fig:LOdiags}
\end{figure}

The basic processes for the production of leptoquark pairs at
the LHC are gluon-gluon fusion and quark-antiquark annihilation:
\begin{eqnarray}
\label{parton-processes}
g + g       \to \LQ + \LQb \nonumber \\
q + \bar{q} \to \LQ + \LQb 
\end{eqnarray}
as shown in Fig.~\ref{fig:LOdiags}. In the pointlike limit, the
gluon-leptoquark interactions are determined by the non-abelian
SU(3)$_{\rm C}$ gauge symmetry of scalar QCD so that the theoretical
predictions for the pair production of scalar leptoquarks are
parameter-free in the first two generations for which Yukawa terms can
safely be neglected~\cite{Grifols:1981aq}:
\begin{eqnarray}
\hat{\sigma}^{\mbox{\scriptsize LO}}_{gg}\,
&=&
\frac{\alpsq\pi}{96\hat{s}}\,
\Big[ \beta \left( 41 - 31 \beta^2 \right)
+ \left( 18 \beta^2 - \beta^4 - 17 \right) 
           \log\frac{1+\beta}{1-\beta} \Big] \nonumber\\[1mm]
\hat{\sigma}^{\mbox{\scriptsize LO}}_{q\bar{q}}\, 
&=&
\frac{\alpsq\pi}{\hat{s}}\,
\frac{2}{27} \, \beta^3 \, ,
\end{eqnarray}
where $\beta=(1-4\msq/\hat s)^{1/2}$ denotes the leptoquark velocity
and ${\hat{s}}^{1/2}$ the invariant energy of the subprocess. As
expected, for low leptoquark masses gluon fusion is by far the
dominant production mechanism. Quark-antiquark annihilation becomes
important for larger masses, providing 30\% of the production cross
section for leptoquark masses of about 1.5 TeV, as will be proved
later in detail.

The calculation of the QCD radiative corrections for the total cross
section of leptoquark production
\begin{eqnarray}
\sigma[pp \to LQ + \overline{LQ}] = \sigma_{gg} + \sigma_{q \bar{q}}
                                    + \sigma_{gq}
\end{eqnarray}
including virtual effects, gluon bremsstrahlung and initial-state
parton splittings at ${\mathcal{O}}(\alpha_{\mathrm{s}})$ has been
described in Ref.~\cite{Kramer:1997hh,Beenakker:1997ut} in detail. In
the following we define the calculational scheme and the physical
input parameters chosen for the present numerical analysis. The
renormalization of the strong coupling $\alpha_{\mathrm{s}}(\mu)$ and
the factorization of initial-state collinear singularities at $\mu$
are performed in the $\overline{\mathrm{MS}}$ scheme. The top quark
and the leptoquark are decoupled from the running of
$\alpha_{\mathrm{s}}(\mu)$. For the calculation of the $pp$ cross
section we have adopted the CTEQ6L1 and CTEQ6M~\cite{Pumplin:2002vw}
parton distribution functions at LO and NLO, corresponding to
$\Lambda_5^{\mathrm{LO}} = 165$~MeV and
$\Lambda_5^{\overline{\mathrm{MS}}} = 226$~MeV at the one- and
two-loop level of the strong coupling $\alpha_{\mathrm{s}} (\mu)$,
respectively.

The QCD corrections strongly affect the parton cross sections near the
production threshold, in particular in $gg$ fusion which rises steeply
with energy. {\it (i)} Sommerfeld rescattering, {\it i.e.} Coulombic
gluon exchange between the final-state partons, is singular near
threshold~\cite{schwinger} in the velocity of the produced
leptoquarks, thus removing the phase-space velocity factor 
$\beta$ from the cross
sections for S-waves and damping the phase-space suppression
${\beta}^3 \to {\beta}^2$ for P-waves.  These corrections are positive
in the attractive color-neutral singlet channels and negative in the
repulsive color-octet channels. For gluon initial states the
color-singlet channels dominate, for quark-antiquark initial states
the color-octet channels. The gluon-exchange ladders can be resummed
\cite{sommerfeld}, giving rise to the threshold correction factors
\cite{Fadin:1990wx} $F_{th}^{1,8} = 1\pm
\frac{1}{2} x_{1,8} \to x_1/(1-e^{-x_1})$ and $x_8/(e^{x_8}-1)$ with
$x_1 = \frac{4}{3} \frac{\pi\alpha_s}{\beta}$ and $x_8 = \frac{1}{6}
\frac{\pi\alpha_s}{\beta}$ for singlet and octet channels, respectively.
{\it (ii)} In addition, large initial-state gluonic bremsstrahlung
corrections of the type $\log^2 \beta$ and $\log \beta$ emerge, where
the universal double logarithm can be exponentiated. The perturbative 
expansion of the total parton cross section can be expressed in terms 
of scaling functions, 
\begin{equation}
\hat{\sigma}_{ij} = 
\frac{\alpsq(\msq)}{\msq} 
\left\{ f_{ij}^{B}(\beta)
 + 4 \pi \alps(\msq) \left[f_{ij}^{V+S}(\beta) + f_{ij}^{H}(\beta) 
\right] \right\}_{[i,j=g,q]} \, , 
\end{equation}
where we define the Born term cross sections of the parton
subprocesses by $f^B$ and the QCD virtual+soft and hard gluon
corrections by $f^{V+S/H}$. The scaling functions depend on the
invariant parton energy ${\hat{s}}^{1/2}$ through $\beta$. The
expressions $f^B$ and $f^{V+S/H}/f^B$ near threshold, $\beta \ll 1$,
\begin{equation}
\begin{array}{ll}
 f_{gg}^B =  \displaystyle{\frac{7 \pi \beta}{384}} & 
 \quad f_{q\bar{q}}^B = \displaystyle{\frac{\pi \beta^3}{54}} \\[3mm]
 f_{gg}^{V+S} / f_{gg}^{B} = \displaystyle{\frac{11}{336 \beta}} &
 \quad f_{q\bar{q}}^{V+S}/f_{q\bar{q}}^{B} = \displaystyle{-\frac{1}{48 \beta}} \\[3mm]
 f_{gg}^{H}  / f_{gg}^{B} =  \displaystyle{\frac{3}{2\pi^2}\log^2(8\beta^2)
  -\frac{183}{28\pi^2} \log(8\beta^2)} & 
 \quad f_{q\bar{q}}^{H} / f_{q\bar{q}}^{B} = \displaystyle{\frac{2}{3\pi^2}\log^2(8\beta^2) 
  - \frac{107}{36\pi^2}\log(8\beta^2)}
\end{array} 
\end{equation}
can be derived from Ref.~\cite{Beenakker:1997ut} by choosing the mass $\mlq$
as renormalization and factorization scale. The scaling functions for
$qg$ initial states are neither linearly nor logarithmically singular
at threshold.
  
For high energies the NLO parton cross sections approach non-zero
limits asymptotically in contrast to the $\sim 1/{\hat s}$ scaling behavior
of the Born cross sections. The asymptotic NLO parton cross sections
read in the nomenclatura of the preceding paragraph:
\begin{equation}
  f_{gg}^H = \frac{2159}{43200\pi}\, , \qquad 
  f_{qg}^H = \frac{2159}{194400\pi} \, .
\end{equation}
The ratio between the gluon-fusion and Compton term is large, $9:2$,
which is a consequence of the large color charges of the gluons
compared to the quarks.
  
\begin{figure}[p]
\includegraphics[width=11.0cm]{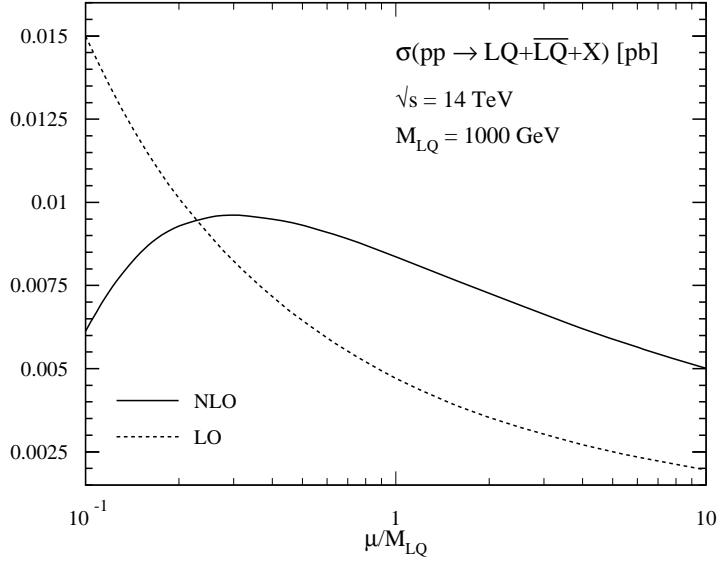}
\caption{Variation of the LO and NLO cross sections for  
 $pp\to{\LQ}+{\LQb}+X$ at the LHC with the renormalization and
 factorization scales. The leptoquark mass has been set to
 $M_{\LQ}=1$~TeV. }
\label{fig:scale}
\end{figure}

The scale dependence of the theoretical prediction is reduced
significantly when higher order QCD corrections are included. This is
demonstrated in Fig.~\ref{fig:scale} where we compare, for an
intermediate leptoquark mass of 1 TeV, the dependence on the common
renormalization/factorization scale $\mu$ at the leading and
next-to-leading order of the total cross section. For the comparison
of the LO and NLO results, we have calculated all quantities [{\it
i.e.} the partonic cross sections, $\alpha_{\mathrm{s}}(\mu)$ and the
parton densities] consistently in leading and next-to-leading
order. The scale dependence of the leading-order cross section is
steep and monotonic. At next-to-leading order the scale dependence is
strongly reduced, and the NLO cross section runs through a broad
maximum near $\mu \approx \mlq/4$ which stabilizes the NLO prediction
very effectively.

\begin{figure}[p]
\includegraphics[width=11.0cm]{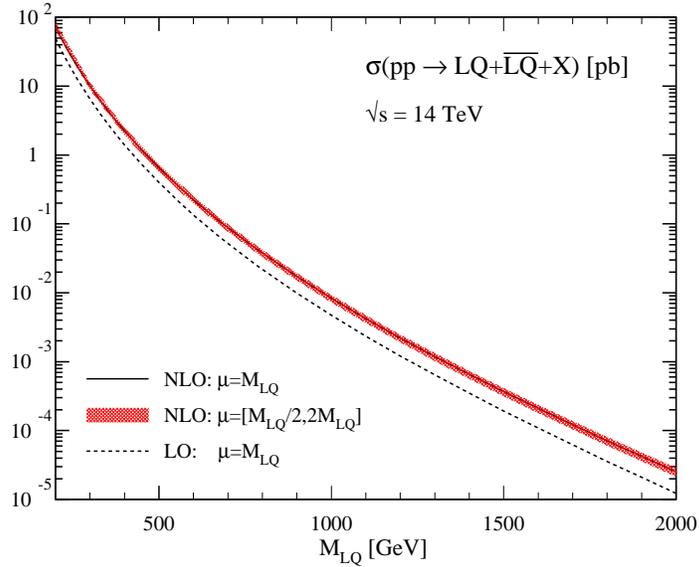}
\caption{Total cross section for $p p \to \LQ + \LQb + X$ at the LHC 
 energy $\sqrt{s} = 14$~TeV as a function of the leptoquark mass
 $M_{\LQ}$. The variation of the NLO cross section with the value of
 the renormalization/factorization scale is indicated by the shaded
 band.}
\label{fig:cxn}
\end{figure}

The QCD radiative corrections enhance the cross section for the
production of leptoquarks in the vicinity of the standard scale choice
$\mu \approx \mlq$, see Fig.~\ref{fig:cxn}. The $K$-factors, $K =
\sigma_{\mbox{\scriptsize NLO}} / \sigma_{\mbox{\scriptsize LO}}$,
with all quantities in the numerator and denominator calculated in NLO
and LO, respectively, are displayed in Table~\ref{table}.  [Note that
$\sigma_{qg}$ is only the remnant of the $qg$ cross section after the
collinear singularities are subtracted via mass factorization;
overcompensation can give rise to negative values and $\sigma_{qg}$
must not be interpreted as a physical cross section.]  The $K$-factors
range from $K \approx 1.5$ at $\mlq \approx 200$~GeV up to $K \approx
1.9$ at the upper end of the LQ mass reach of $\approx 1500$~GeV at
LHC. The leptoquark results coincide with the cross sections for the
production of squark--antisquark pairs in the limit of large gluino
mass~\cite{Beenakker:1994an,Beenakker:1996ch,Beenakker:1997ut} after
the small ($< 0.5\%$) contributions from rescattering diagrams
involving four-squark self-couplings are eliminated. This comparison
provides us with an independent check of the present numerical
analysis.  [Agreement with the scalar cross sections for low-mass
leptoquarks in the compilation of Ref.~\cite{Blumlein:1998ym} is only
of limited value as the $K$-factors presented in
Ref.~\cite{Blumlein:1998ym} have been obtained by using the NLO code
of the calculation Ref.~\cite{Kramer:1997hh}.]  Despite the sizable
corrections at the scale $\mu \approx \mlq$, the moderate scale
dependence of the NLO result implies a reliable perturbative
expansion.\footnote{For the production of Higgs bosons in gluon-gluon
  fusion, NLO $K$-factors have been found in the range between 1.5 and
  2, cf.~Ref.~\cite{Graudenz:1992pv}, but they are modified only by
  another 20\% in NNLO~\cite{Harlander:2002wh}.  From this example we
  may infer that also the isomorphic perturbative QCD expansion in
  leptoquark production is expected to converge sufficiently fast and
  that the NLO results are of physical significance.}

To study the uncertainty in the prediction of the cross section
inferred by the parametrization of the parton densities, we have
repeated the calculation using the MRTS2002
parametrization~\cite{Martin:2001es}. The difference between the NLO
results based on MRST2002 and CTEQ6, displayed in Table~\ref{table},
increases with increasing leptoquark mass due to the uncertainty in
the gluon distribution at large $x$, but does not exceed 10\% in the
experimentally accessible mass range up to $\mlq \simeq$ 1.5 TeV.

{\it In conclusion.} As anticipated from many calculations performed
in the past, the NLO analysis stabilizes the theoretical prediction of
the production cross section for leptoquarks. Moreover, for the
standard choice of the renormalization and factorization scales near
the leptoquark mass, the higher order corrections increase the cross
section so that Born calculations, performed for the same scale of QCD
coupling and parton densities, provide us with a conservative lower
limit. By the same token, the NLO corrections shift the LQ mass limit
upward by an amount of about 100 GeV at the upper end of the LQ mass
spectrum that can be probed at LHC.

\begin{acknowledgments}
We thank Simona Rolli for discussions on leptoquark production at the
hadron colliders Tevatron and LHC that have initiated this study at
the FNAL Tev4LHC Workshop. P.M.~Zerwas thanks the Fermilab Theory Group
for the warm hospitality during an extended visit.
\end{acknowledgments}

\vspace*{2cm}

\clearpage

\begin{table}[p]
\begin{center}
\caption{\label{table} Total cross section and $K$-factors for 
  $pp\to\LQ+\LQb+X$ at the LHC energy $\sqrt{S} = 14$~TeV. The
  renormalization and factorization scales have been set to
  $\mu=\mlq$. The CTEQ6L1 and CTEQ6M~\cite{Pumplin:2002vw} parton
  densities have been adopted as default, whereas in the rightmost
  column a comparison is performed with the
  MRST2002~\cite{Martin:2001es} parton densities. The relative weight
  of $gg$, $q\bar{q}$ and $gq$ contributions to the cross section at
  NLO is given in the second-to-last column. [The negative sign of
  ${\sigma}_{gq}$ is a mere artifact of subtracting collinear
  initial-state singularities via mass factorization; by definition,
  this particular higher-order quantity must not be interpreted as a
  cross section.]}
\vspace*{5mm}
\begin{tabular}{|c||c|c|c|c||c|}
\hline
 & \multicolumn{4}{c||}{\small CTEQ6(LO/NLO)} &
 \rule[-3mm]{0mm}{8mm} {\small MRST2002} \\ \cline{2-6}
 \raisebox{1.9ex}[-1.9ex]{ $\mlq$~[GeV]} &
 $\sigma_{\rm LO}$~[fb] & $\sigma_{\rm NLO}$~[fb] & 
 $gg : q\bar{q} : gq$   & K & \rule[-3mm]{0mm}{8mm} $\sigma_{\rm NLO}$~[fb] \\
 \hline\hline
  200  & $0.500 \times 10^{2}$  & $0.742 \times 10^{2}$  & $0.94  : 0.05  : \;\;\; 0.01$  & 1.48 & $0.779 \times 10^{2}$  \\
  400  & $0.140 \times 10^{1}$  & $0.224 \times 10^{1}$  & $0.91  : 0.10  : -0.01$  & 1.60 & $0.243 \times 10^{1}$  \\
  600  &  0.135                 &  0.225                 & $0.88  : 0.15  : -0.03$  & 1.67 &  0.245                 \\
  800  & $0.219 \times 10^{-1}$ & $0.378 \times 10^{-1}$ & $0.84  : 0.19  : -0.03$  & 1.73 & $0.406 \times 10^{-1}$ \\
  1000 & $0.471 \times 10^{-2}$ & $0.836 \times 10^{-2}$ & $0.82  : 0.22  : -0.04$  & 1.77 & $0.879 \times 10^{-2}$ \\
  1200 & $0.121 \times 10^{-2}$ & $0.221 \times 10^{-2}$ & $0.81  : 0.24  : -0.05$  & 1.83 & $0.226 \times 10^{-2}$ \\
  1400 & $0.349 \times 10^{-3}$ & $0.655 \times 10^{-3}$ & $0.79  : 0.26  : -0.05$  & 1.88 & $0.650 \times 10^{-3}$ \\
  1600 & $0.109 \times 10^{-3}$ & $0.210 \times 10^{-3}$ & $0.78  : 0.28  : -0.06$  & 1.93 & $0.201 \times 10^{-3}$ \\
  1800 & $0.357 \times 10^{-4}$ & $0.713 \times 10^{-4}$ & $0.77  : 0.29  : -0.06$  & 2.00 & $0.656 \times 10^{-4}$ \\
  2000 & $0.122 \times 10^{-4}$ & $0.253 \times 10^{-4}$ & $0.77  : 0.30  : -0.07$  & 2.07 & $0.222 \times 10^{-4}$ \\
\hline
\end{tabular}
\end{center}
\end{table}

\end{document}